\documentclass[runningheads]{llncs}
\pdfoutput=1
\usepackage{graphicx}
\usepackage{amssymb}
\usepackage{booktabs}
\usepackage{array}
\usepackage{amsmath}
\newcolumntype{N}{>{\centering\arraybackslash}m{.5in}}
\newcolumntype{G}{>{\centering\arraybackslash}m{2in}}
\usepackage{varwidth}
\usepackage{caption}
\usepackage{subcaption}
\usepackage[misc]{ifsym}
\usepackage{bbding}

\begin{document}

\title{Unified Embeddings of Structural and Functional Connectome via a Function-Constrained Structural Graph Variational Auto-Encoder}
\titlerunning{Unified structural-functional joint embedding with FCS-GVAE}

 \institute{Department of Biomedical Engineering, University of Illinois Chicago, Chicago, IL, USA \\ \email{camode2@uic.edu}\\ \and Department of Psychiatry, University of Illinois Chicago, Chicago, IL, USA \and Department of Information and Decision Sciences, University of Illinois Chicago, Chicago, IL, USA \and Department of Electrical and Computer Engineering, University of Pittsburgh, Pittsburgh, PA, USA }

\maketitle

\begin{abstract} 
Graph theoretical analyses have become standard tools in modeling functional and anatomical connectivity in the brain. With the advent of connectomics, the primary graphs or networks of interest are structural connectome (derived from DTI tractography) and functional connectome (derived from resting-state fMRI). However, most published connectome studies have focused on either structural or functional connectome, yet complementary information between them, when available in the same dataset, can be jointly leveraged to improve our understanding of the brain. To this end, we propose a function-constrained structural graph variational autoencoder (FCS-GVAE) capable of incorporating information from both functional and structural connectome in an unsupervised fashion. This leads to a joint low-dimensional embedding that establishes a unified spatial coordinate system for comparing across different subjects. We evaluate our approach using the publicly available OASIS-3 Alzheimer’s disease (AD) dataset and show that a variational formulation is necessary to optimally encode functional brain dynamics. Further, the proposed joint embedding approach can more accurately distinguish different patient sub-populations than approaches that do not use complementary connectome information.

\end{abstract}

\section{Introduction}\label{sec:introduction}

Advances in magnetic resonance imaging (MRI) technology have made very large amounts of multi-modal brain imaging data available, providing us with unparalleled opportunities to investigate the structure and function of the human brain. Functional magnetic resonance imaging (fMRI), for example, can be used to study the functional activation patterns of the brain based on cerebral blood flow and the Blood Oxygen Level Dependent (BOLD) response \cite{fox2007spontaneous}, whereas diffusion tensor imaging (DTI) can be used to examine the \emph{wiring diagram} of the white matter fiber pathways, i.e., the structural connectivity of the brain \cite{assaf2008diffusion}.

Because of their utility in understanding human brain structure, in examining neurological illnesses, and in developing therapeutic/diagnostic applications, brain networks (also called \emph{connectomes}) have attracted a lot of attention recently. The principal networks of interest are structural brain networks (derived from DTI) and functional brain networks (derived from fMRI). Graph-based geometric machine learning approaches have shown promise in processing these connectomics datasets due to their ability to leverage the inherent geometry of such data. However the majority of these existing studies in brain network analysis tend to concentrate on either structural or functional connectomes \cite{sporns2013structure} \cite{van2010exploring}. Our hypothesis is that both the anatomical characteristics captured by structural connectivity and the physiological dynamics properties that form the basis of functional connectivity can lead to a much improved understanding of the brain's integrated organization, and thus it would be advantageous if both structural and functional networks could be analyzed simultaneously. 

To this end, in this study we propose employing a graph variational autoencoder (GVAE) \cite{Kipf2016} based system to learn low-dimensional embeddings of a collection of brain networks, which jointly considers both the structural and functional information. Specifically, we employ graph convolutions to learn structural and functional joint embeddings, where the graph structure is defined by the structural connectivity and node properties are determined by functional connectivity. The goal here is to capture structural and functional network-level relationships between subjects in a low-dimensional continuous vector space so that inferences about their individual differences, as well as about the underlying brain dynamics, can be made. Experimental results in Section~\ref{sec:experiments} show how the embedding space obtained through this preliminary line of research enables comparison of higher-order brain relationships between subjects. We also validate the usefulness of the resulting embedding space via a classification task that seek to predict whether the subject is affected by AD or not, and show how it outperforms single modality baselines.

\section{Proposed Framework}\label{sec:framework}

\subsection{Problem Statement}
We are given a set of brain network instances \(D={{G_1,...,G_M}}\), where each instance \(G_i\) corresponds to a different subject and is composed of a structural \(G_i^{(s)}\) and a functional \(G_i^{(f)}\) network. Our goal is to jointly embed the information contained in the two networks in order to obtain a common coordinate space that enables interpretable comparisons between subjects. We will test the quality of the embeddings in this common space via a downstream task that involves classifying AD subjects and healthy subjects.

Note that most existing approaches focus on one of the network types (structural or functional) to learn embeddings. Instead, our objective is to design an unsupervised learning approach that can extract the complimentary information that exists across these two modalities to improve the quality of embeddings, and thus the ability to improve downstream tasks.

\subsection{Our Modeling Framework}

\begin{figure}[h]
\begin{center}
\includegraphics[scale=0.3]{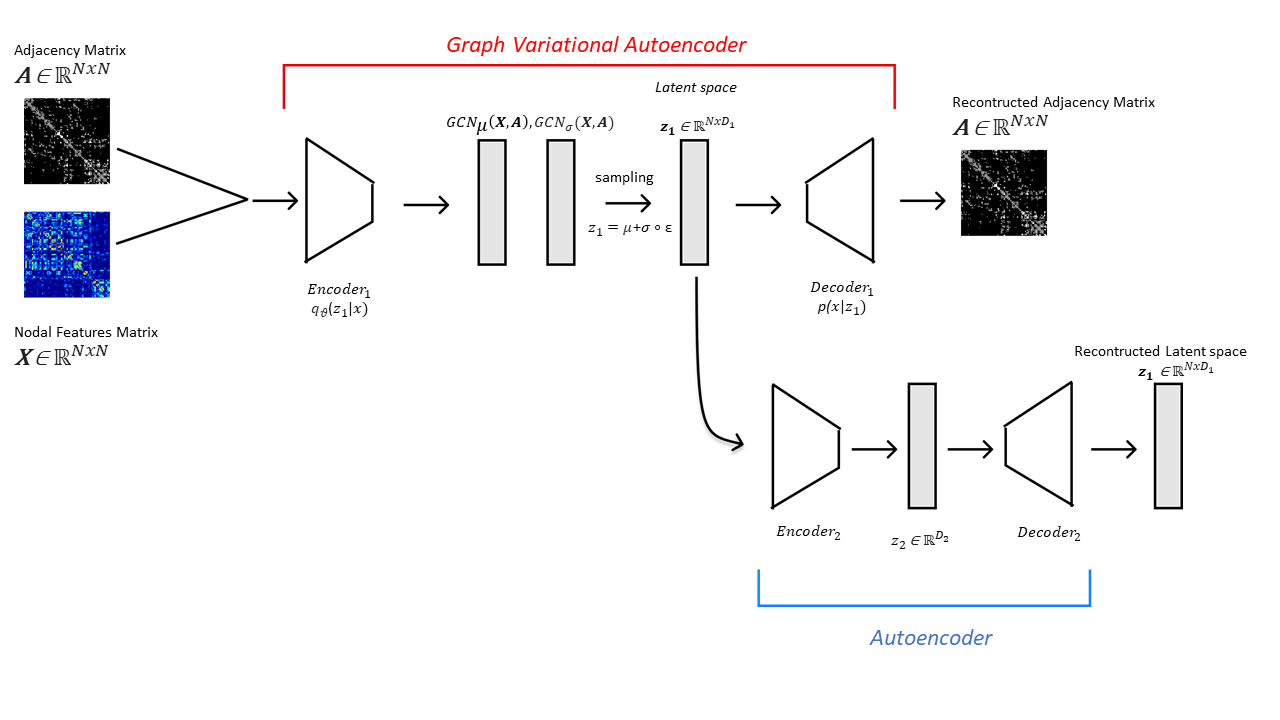}
\caption{An overview of the proposed FCS-GVAE model: The adjacency matrix \(A\) used as input is represented by the structural connectivity network. An \textit{n}-dimensional feature vector \(\mathbf{x_i}\) is assigned on each node, which is the vector of the corresponding node in the functional connectivity network. The resulting encoded data, \(\mathbf{Z}_1\) is then further compressed using an autoencoder, whose latent vector, \(\mathbf{Z_2}\), is the one used to compare and contrast different subjects.}
\label{fig:workflow}
\end {center}
\end{figure}

We propose an augmented function-constrained structural graph variational autoencoder based system, or FCS-GVAE, that involves a GVAE and an Autoencoder (AE), as shown in Figure \ref{fig:workflow}. We employ graph convolutions to learn the structural and functional joint embeddings, \(\mathbf{Z_1} \in \mathbb{R}^{N\times D_1}\), where \(D_1\) represents the dimensionality of each node embedding. The sampling at the GVAE bottleneck is similar to the one of a traditional VAE, with straightforward Gaussian sampling (and the only assumption taken is the one of gaussianity of the latent variables). Given the joint node level embedding matrix, a graph-level embedding \(\mathbf{Z_2} \in \mathbb{R}^{D_2}\) is then obtained through an AE, where \(D_2\) is the chosen dimensionality of the graph-level embedding. These resultant embeddings, one per subject, are then visualized via t-distributed stochastic neighbor (\textit{t-sne}) \cite{van2008visualizing}.

Once the parameters of the GVAE and AE are learned, each subject's data is transformed into the structural functional joint embedding as follows. The structural connectivity, SC, is the adjacency input matrix, \(A \in \mathbb{R}^{N\times N}\), and the functional connectivity, FC, is the matrix of nodal features, \(X \in \mathbb{R}^{N\times C}\), where \(C=N\). In Section~\ref{sec:experiments}, we compare the value of designing our embedding architecture this way against a baseline that removes the FC nodal feature data, as suggested in \cite{li2019mapping}. The quality of these embeddings is assessed using a downstream classification task involving the detection of AD in subjects.
Next, we briefly discuss the notion of graph convolutions and GVAE and justify the need for GVAE vs a regular graph auto-encoder as well as the need for another downstream AE that transforms node embeddings to a graph embedding.

\subsection{Graph Convolution}

A graph convolutional layer~\cite{Kipf2017} works as follows. Let \(G = \{\mathbb{V},A\}\) be a weighted, undirected graph, where \(\mathbb{V}\) is a set of \(N\) nodes, and \(A \in \mathbb{R}^{N\times N}\) is an adjacent matrix, specifying the inter-nodal connections. The normalized Laplacian is defined as \(L=I_N - D^{-1/2} A D^{-1/2}\), with \(I_N\) representing the \(N\)-dimensional identity matrix and the diagonal degree matrix \(D\) having entries \(D_{i,i}= \sum_{j} A_{i,j}\). \(L\) can be further decomposed into the form \(U \Lambda U^T\), where \(U\) is the matrix of eigenvectors of \(L\) and \(\Lambda\) is the diagonal matrix of its eigenvalues.

Let \(\mathbf{x}\) be the attribute vector corresponding to a node. 

Given a filter \(\mathbf{h} \in \mathbb{R}^N\), the graph convolutional operation between  \(\mathbf{x}\) and a filter \(\mathbf{h}\) is defined as: 

\begin{equation}
\label{eq:graph convolution}
 \mathbf{x} \circledast \mathbf{h} = U((U^T \mathbf{h}) \odot (U^T \mathbf{x}))=U{\text{\^H}}U^T \mathbf{x},
 \end{equation}
 
 where \({\text{\^H}}=\mathrm{diag}(\mathbf{\theta})\) replaces \(U^T \mathbf{h}\) with \(\boldsymbol{\theta} \in \mathbb{R} ^N\) parameterizing it in the Fourier domain.
 Since the evaluation of Equation \ref{eq:graph convolution} requires explicit computation of the eigenvector matrix, it can be prohibitively costly for very large networks; consequently, \({{\text{\^H}}}\) has been approximated through a truncated expansion in terms of the Chebyshev polynomials \cite{Defferrard2016,Hammond2011}.
 \ Kipf et al. \cite{Kipf2017} subsequently provided a second order approximation, such that 
 \(\mathbf{x} \circledast \mathbf{h} \approx \theta(I_n + D^{-1/2} A D^{-1/2})\mathbf{x}\).
 Generalizing the \(\mathbf{x}\) vector to the matrix \({X} \in \mathbb{R}^{N\times C}\) with C input channels,  
the following graph convolutional layer filtering is introduced: 
\begin{equation}\label{eq:GCN}
Z={\tilde{A} X \tilde{\Theta}}
\end{equation}
With \(\tilde{A} = I_n + D^{-1/2} A D^{-1/2} \), \(\tilde{\Theta} \in \mathbb{R}^{C\times F}\) a matrix of the \(F\) filter parameters to be learned, and \(Z \in \mathbb{R}^{N\times F}\) the convolved signal matrix. 
Stacking numerous convolutional layers (see Equation \ref{eq:GCN} for instance), each of them followed by the application of a non-linearity \cite{wu2020comprehensive}, graph convolutional networks are defined as:

\begin{equation}\label{eq:GCN_layer}
GCN^{l+1}(\tilde{A},X)=\sigma (\tilde{A}\,      GCN^{l}(\tilde{A},X)^{l}\, \tilde{\Theta}_{l+1}),
\end{equation}
where \(\sigma(\cdot)\) represents the activation function 
and \(GCN^{0}(\tilde{A},X) :=X\). Important to our setting, note that equations \ref{eq:GCN} and \ref{eq:GCN_layer} show how the output of the graph convolution process contains not only the network's topological information (represented by $A$), but also the nodal properties (represented by $X$). Next we discuss the GVAE component of our setup.
  
\subsection{Graph Variational Autoencoder}

Variational Autoencoder (VAE)~\cite{kingma2013auto} is a variant of deep generative models used for learning latent representations from an unlabeled dataset by simultaneously training an encoder and a decoder to maximize the evidence lower bound (ELBO). The encoder maps the data into a low-dimensional latent representation \(z\). The \(z\) is then sampled from the approximate posterior distribution \(q(z|X)\), typically chosen to be an independent Gaussian distribution \(\mathcal{N}(\mu , \mathrm{diag}(\sigma^2))\), where \(\mu\) and \(\sigma \) are output by the encoder. The decoder reconstructs the original data by deriving the likelihood of data \(X\) based on the variable \(z\), \(p(X|z)\).

GVAE extends this idea to graph-structured data, and in learning the latent representation of the data on which it is trained, it incorporates both the input network's topological characteristics, as defined by the adjacency matrix, and the network's node features \cite{Kipf2016}. In a GVAE, the encoder is parameterized by a series of graph convolutional network layers. Technical details about the approximate posterior distribution, as well as the justification for using a non-trainable innerproduct decoder can be found in \cite{Kipf2016} and \cite{kingma2013auto}.

\subsection{Justifying the Choice of GVAE and AE Components}\label{sec:justification}

As shown in Figure \ref{fig:workflow}, our model involves a cascade of two auto-encoders, one of which is generative. We need the graph auto-encoder to be generative because we are not only interested in compression but also the generalizability of the encoders and decoders. For instance, we can reuse the decoder to generate/sample new structural connectomes, which is not possible with a vanilla auto-encoder. Further, since the GVAE produces high-granularity node-level embeddings, we use a straight-forward AE to compress that information to a low enough dimension for downstream visualization/supervised learning tasks. While doing so, we are able to obtain a unified graph level embedding (for a given structural functional connectome pair) that can be much better than naive approaches (such as averaging node-embeddings). Note that this second component, the AE, may be replaced, for example by a PCA. However, unlike PCA, a single layer AE with no-nonlinearity does not impose orthogonality, hence reducing the reconstruction error. Nonetheless, in our case we make use of multiple linear layers followed by appropriate non-linearities to achieve maximal compression (which is beyond what PCA can achieve). As can be seen in the following section, these choices indeed show the value of complementary information present in both networks and how they help define better embeddings (as evaluated using the AD classification task).

\section{Experimental Evaluation and Results} \label{sec:experiments}

We discuss the dataset, the hyperparameter choices, and how resulting embeddings improve on existing methods, using both structural and functional network information simultaneously. To evaluate our node and graph embeddings, we use an auxiliary link prediction problem and the AD classification problem respectively. Both these tasks quantitatively show the value of our joint embedding approach compared to natural baselines.

\subsection{Dataset}
MRI imaging used in this study comes from the OASIS-3 dataset for Normal Aging and Alzheimer’s disease \cite{lamontagne2019oasis} and was collected in a 16-channel head coil of a Siemens TIM Trio 3T scanner.  OASIS-3 includes the clinical, neuropsychological, neuroimaging, and biomarker data of 1098 participants (age: 42–95 years; www.oasis-brains.org). We analyzed the data from 865 participants with combined structural and functional MRI sessions (N = 1326). The dataset includes 738 Females (112 with AD), and 588 Males (163 with AD). AD has been defined here as having a clinical dementia rating greater than 0.

The brain regions considered for this study, which cover the entire brain, consist of $132$ regions: $91$ cortical Region of Interests (ROIs) are obtained from the FSL Harvard-Oxford Atlas maximum likelihood cortical atlas, $15$ subcortical ROIs are obtained from the FSL Harvard-Oxford Atlas maximum likelihood subcortical atlas \cite{desikan2006automated}, and the remaining $26$ are cerebellar ROIs from the AAL atlas \cite{tzourio2002automated}. 

The brain structure connectivity graph is generated from combining brain grey matter parcellation extracted from T1-weighted MRI and the white matter fiber tracking obtained from DTI acquisition. The graph is undirected and each node \(v_i\) in \(V\) denotes a specific brain region of interest (ROI). Element \({A}_{ij}\)
in \(A\), the adjacency matrix, denotes the weight of the connection between the two nodes \(v_i\) and \(v_j\).  
Note that we performed minimal processing on the structural graph and have generated functional graphs via Pearson coefficients of BOLD signals in a particular time window (without thresholding). Thus, the choice of time windows is the only preprocessing step.

\subsection{Hyperparameter Choices}

We chose $L = 4$ graph convolutional layers to model the encoder of the GVAE component. The number of layers for GVAE is determined by computing the diameter of the adjacency matrix, defined as \(\max_{v_i,v_j}(\textrm{dist}(v_i,v_j)), \forall{v_i,v_j} \in V\), where \(\textrm{dist}(v_i,v_j)\) represents the shortest path to reach \(v_j\) from \(v_i\). The maximum graph diameter for our dataset turned out to be $4$, implying that every node reaches/influences every other node's representation after $4$ graph convolution layers. Note that this choice is motivated by the need to avoid the \textit{vanishing gradient}  phenomena~\cite{li2018deeper}.

\begin{figure}
     \centering
     \begin{subfigure}[h]{0.3\textwidth}
         \centering
         \includegraphics[width=\textwidth]{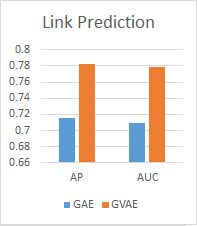}
         \caption{}
         \label{fig:GAE_GVAE}
     \end{subfigure}
     \begin{subfigure}[h]{0.6\textwidth}
         \centering
         \includegraphics[width=\textwidth]{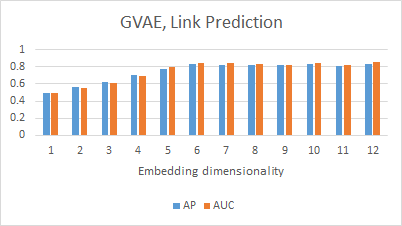}
         \caption{}
         \label{fig:Link_Pred}
     \end{subfigure}
     \hfill
        \caption{ Performances of the two different models on the link prediction task: a) FCS-GVAE performs better than the GAE for both the considered metrics. b) Performances obtained by varying the dimensionality of the embedding layer. It can be seen that a plateau is reached when the dimensionality is $\geq 6$.}
        \label{fig:Justifications}
\end{figure}

The dimension \(D_1\) of the GVAE latent space is selected by taking in consideration the Average Precision (AP) and the Area Under the Curve (AUC) of an auxiliary link prediction problem that is solved while maximizing ELBO.
Based on Figure \ref{fig:Link_Pred}, we set the \(D_1\) parameter to be equal to $6$. Note that this result also suggests that the non-linear dynamics of the brain (as defined using structural and functional graphs) can essentially be captured by a $6$ dimensional node embedding space. In addition to the methodological justification for choosing GVAE compare to a vanilla graph auto-encoder given in Section~\ref{sec:justification}, we also show empirically that GVAE performs much better when compared to a vanilla/non-variational graph auto-encoder (GAE) on the link prediction task (see Figure~\ref{fig:GAE_GVAE}). The first graph convolutional layer in the encoder of the GVAE component is characterized by $48$ filters, the second one by $24$, the third by $12$ and both \(GCN_{\mu}^{4}(A,X)\) and \(GCN_{\sigma}^{4} (A,X)\) have $6$ filters each. To maximize evidence lower bound (ELBO), we use a gradient descent variant known as Adaptive Moment Estimation (ADAM)~\cite{Kingma2015}, setting a learning rate equal to $0.001$. 

We further compress the latent matrices corresponding to node-level embeddings of a subject, viz., \(\mathbf{Z_1} \in \mathbb{R}^{132\times 6}\), into a smaller subject level graph embedding vector, \(z_2 \in \mathbb{R}^{D_2}\) through the vanilla AE component (see Figure~\ref{fig:workflow}). This second latent space is obtained through an encoding structure composed of 5 linear layers, with each of them using the ReLU activation function. To learn the parameters of the AE, we employ a mean squared error (MSE) loss function, which we minimize using the same ADAM optimizer mentioned earlier. The learning rate is set to $0.001$ and the weight decay is set to $0.00005$. The dimensionality \(D_2\) is set to the value $6$ based on the convergence of the MSE loss.

\subsection{Results Demonstrating the Quality of Joint Embeddings}

\begin{figure}[h]
\begin{center}
\includegraphics[scale=0.4]{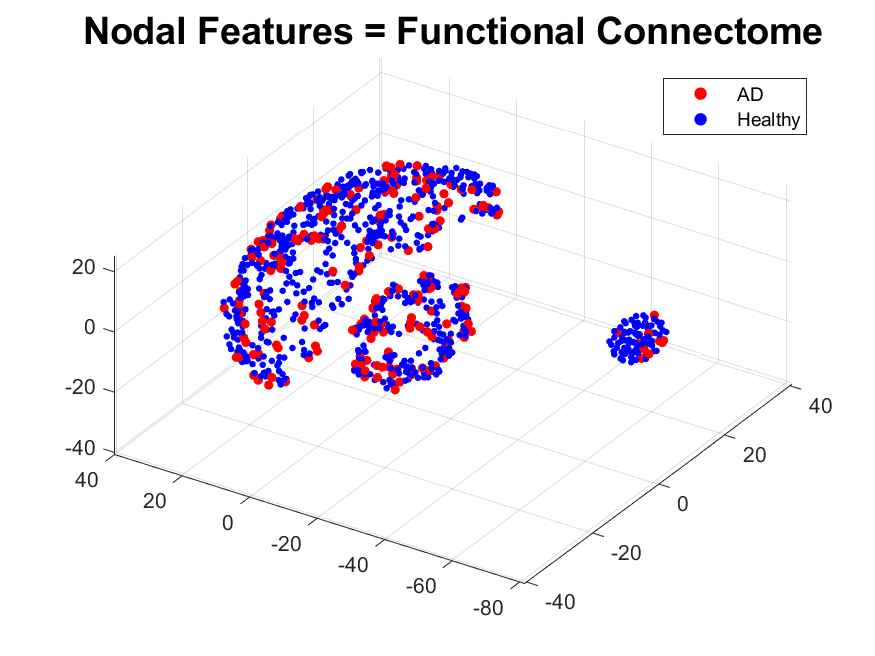}
\includegraphics[scale=0.4]{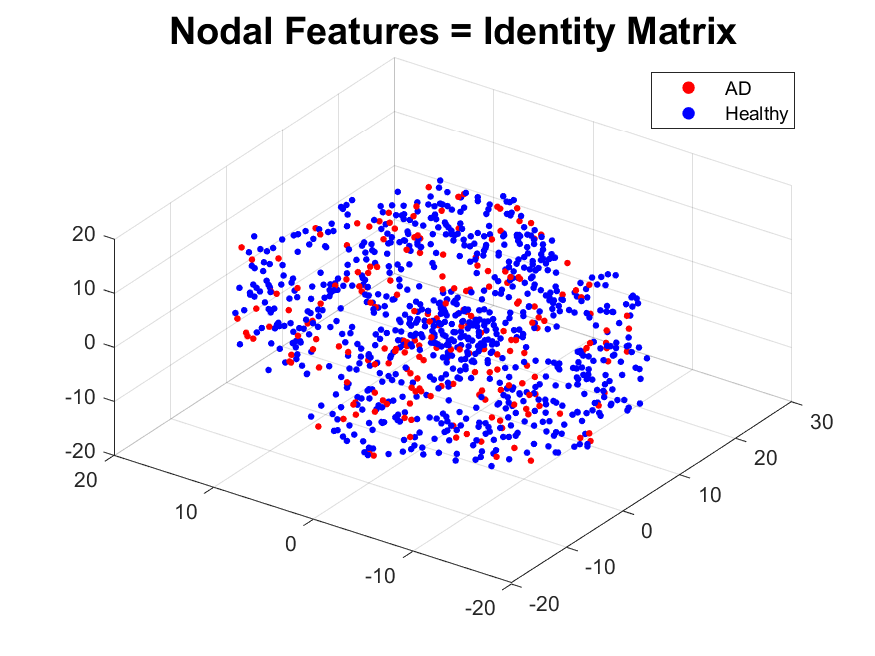}

\caption{T-sne projections of the learned embeddings. When using the joint embeddings computed by FCS-GVAE, different sub-populations are captured.}

\label{embedd}
\end{center}
\end{figure}

A key point of our work is to identify how including the functional connectome (as nodal features) can help further characterize the underlying biology, relative to a baseline model that does not use it (instead using the identity matrix as the nodal feature, a common practice in the graph neural networks research community \cite{li2019mapping}).  
Indeed, Figure \ref{embedd} qualitatively shows how including the FC (left panel) leads to better clustering in the latent space with respect to diagnostic labels (e.g., the right cluster comprises primarily healthy subjects), likely by forcing the latent embedding to account for the temporal dynamics of FC. By contrast, right panel shows the embedding with structural information alone.

We also evaluate the performance of our learned embeddings for AD detection. Here, the latent embedding vector representations across all brain regions of an individual are used as features to predict their diagnostic label (AD versus not AD) using $2$ classic classification models: multi-layer perceptron and random forest. With the hyperparameters set to default values in sklearn, we ran classification by randomly splitting the data into training/test sets (80\%/20\%), and reported performance over the test set with cross-validation. The results (Table \ref{Classification})
highlight the value of using functional connectome as the nodal features, as evidenced by the performance improvement over the baseline model (identity matrix as the nodal feature) in terms of F1-score, precision, and recall.

\begin{table}[h!]
\caption{Classification performances under two different embedding approaches. Using the FC as nodal features (right column) leads to an overall improvement in the classifier performance when compared to simply using the identity matrix as nodal features (left column).}
\centering
\resizebox{.7\textwidth}{!}{%
\begin{tabular}{m{1cm}|c|c|c|c|c|c|}
\cline{2-7}
     & \multicolumn{3}{c|}{\textbf{Identity Matrix}} & \multicolumn{3}{c|}{\textbf{Functional Connectome}} \\  
      \hline
\multicolumn{1}{|c|}{\textbf{Model}} & \textbf{Precision} & \textbf{Recall} & \textbf{F1-score} & \textbf{Precision} & \textbf{Recall} & \textbf{F1-score} \\ \hline

\multicolumn{1}{|c|}{Multi Layer Perceptron} & 0.308$\pm${0.018} & 0.045$\pm${0.012} & 0.109$\pm${0.031} & 0.587$\pm${0.014} & 0.762$\pm${0.045} & 0.663$\pm${0.023} \\ \hline
\multicolumn{1}{|c|}{Random Forest} & 0.304$\pm${0.013} & 0.189$\pm${0.042} & 0.233$\pm${0.032} & 0.573$\pm${0.009} & 0.706$\pm${0.002} & 0.63$\pm${0.014}\\ \hline

\end{tabular}
}%
\label{Classification}
\end{table}

\section{Conclusion}

 We introduced a variational graph auto-encoder framework to unify DTI structural and resting-state functional brain networks, allowing for the definition of a common-coordinate embedding space, i.e., a joint structure-function representation of the brain. When trained on a large AD dataset, this joint embedding framework was able to uncover biologically meaningful sub-populations in an unsupervised manner. Further, we quantitatively demonstrated improvement in classification tasks with our variational formulation (versus a more traditional non-variational graph auto-encoder), suggesting that a variational framework is necessary in optimally capturing functional brain dynamics. In the future, we will aim to further use our approach to uncover the biological underpinnings of different AD sub-populations.\\ \\
 \textbf{Acknowledgement}. This study is partially supported by the NIH (R01AG071243 and R01MH125928) and NSF (IIS 2045848).

\bibliographystyle{plain}
\bibliography{Bibliography.bib}
\end{document}